\documentclass[12pt,a4]{revtex4}
\usepackage[british]{babel}
\usepackage{amssymb,amsmath,amsthm, bbm, mathrsfs}
\usepackage[T1]{fontenc}
\usepackage{latexsym} 
\usepackage{times}
\usepackage{color}
\usepackage{xcolor}
\usepackage{graphicx}
\usepackage[T1]{fontenc}
\usepackage{comment}
\usepackage{t-angles}
\usepackage[framemethod=tikz]{mdframed}
\definecolor{refkey}{rgb}{0,0,1}
\definecolor{labelkey}{rgb}{1,0,0}
\usepackage{cleveref}
\colorlet{shadecolor}{gray!20}

\definecolor{ag}{rgb}{0.29, 0.33, 0.13}
\definecolor{darkblue}{rgb}{0.0, 0.0, 0.55}
\definecolor{darkcerulean}{rgb}{0.03, 0.27, 0.49}
\definecolor{darkpowderblue}{rgb}{0.0, 0.2, 0.6}
\definecolor{britishracinggreen}{rgb}{0.0, 0.26, 0.15}
\newenvironment{blu}{\color{darkpowderblue}}{}
\newenvironment{mgg}{\color{magenta}}{}

\newcommand{\bblu}{\begin{blu}}
\newcommand{\eblu}{\end{blu}}
\newcommand{\bmag}{\begin{mgg}}
\newcommand{\emag}{\end{mgg}}
\newcommand{\bas}{\begin{mgg}}
\newcommand{\eas}{\end{mgg}}
\definecolor{refkey}{rgb}{0,0,1}
\definecolor{labelkey}{rgb}{1,0,0}
\def\<{{\langle}} 
\def\>{{\rangle}}

\def\note#1{{}}

\def\note#1{} 
\def\beq{\begin{equation}} 
\def\eeq{\end{equation}}

\newcounter{zlist} 

\newcounter{blist} 
 
\newcounter{rlist} 
 
 \def\stac#1{\raise-.2cm\hbox{$\stackrel{\displaystyle\otimes}{\scriptscriptstyle{#1}}$}}

\def\cten#1{\raise-.2cm\hbox{$\stackrel{\displaystyle\widehat{\otimes}}{\scriptscriptstyle{#1}}$}}

\def\Label#1{\label{#1}\ifmmode\llap{[#1] }\else 
\marginpar{\smash{\hbox{\tiny [#1]}}}\fi} 
\def\Label{\label} 
\theoremstyle{definition}

\theoremstyle{remark}

\newcounter{c} 
 
\newcommand{\etyk}[1]{\vspace{-7.4mm}$$\begin{equation}\Label{#1} 
\addtocounter{c}{1}} 
\renewcommand{\]}{\ifnum \value{c}=1 $$\else \end{equation}\fi} 
\def\CC{{\mathbb C}}
\newcommand{\Cc}{\mathcal{C}}

\def\*C{{}^*\hspace*{-1pt}{\Cc}}
\def\text#1{{\rm {\rm #1}}}

\def\1{\mathbf{1}}

\newcounter{mnotecount}[section]
\renewcommand{\themnotecount}{\thesection.\arabic{mnotecount}}
\newcommand{\mnote}[1]%{}
{\protect{\stepcounter{mnotecount}}$^{\mbox{\footnotesize
$%\!\!\!\!\!\!\,
\bullet$\themnotecount}}$ \marginpar{%\color{red}%
\raggedright \tiny\em
$\bullet$\themnotecount: #1} }
\numberwithin{equation}{section}
\usepackage{pgfplots}
\pgfplotsset{width=8cm,compat=1.10}
\definecolor{dark_green}{rgb}{0.0, 0.5, 0.0}
\usepgfplotslibrary{fillbetween}
\begin{document} 
\vspace*{-2cm}
\title{On stability of Friedmann-Lema{\^i}tre-Robertson-Walker solutions in doubled geometries.} 
\author{Arkadiusz Bochniak}
\author{Andrzej Sitarz}
%\thanks{Authors acknowledge support by NCN grant OPUS 2016/21/B/ST1/02438}
\affiliation{Institute of Theoretical Physics, Jagiellonian University,
\hbox{prof.\ Stanis\l awa \L ojasiewicza 11, 30-348 Krak\'ow, Poland.}}
%%%%%%%%%%%%%%%%%%
%%\pacs{23.23.+x, 56.65.Dy}
%%%%%%%%%%%%%%%%%%%%%%%%%%%%%%%%%%%%%%%%%%%%%%%%%%%%  
\begin{abstract} 
Motivated by the models of geometry with discrete spaces as additional dimensions we 
investigate the stability of cosmological solutions in models with two metrics of the
Friedmann-Lema{\^i}tre-Robertson-Walker type. We propose an effective gravity action 
that couples the two metrics in a similar manner as in bimetric theory of gravity and 
analyse whether standard solutions with identical metrics are stable under small 
perturbations. 	
\end{abstract} 
\maketitle 
 \vspace*{-1cm}
%%%%%%%%%%%%%%%%%%%%%%%%%%%%%%%%%%%%%%%%%%%%%%%%%%%%  
\section{Introduction}

The spectacular success of geometry in the description of large-scale structure of the Universe (general relativity) as well as fundamental interactions
(gauge theories) is one of the biggest achievements of modern physics. Yet the link between these two is still a major challenge to our understanding
of the world. Apart from that there are multiple efforts to solve the puzzle of dark matter with interesting attempt to modify gravity. The bimetric 
theory \cite{hr1}, being one of consistently formulated models, appears to be a good candidate to solve the puzzle in accordance with the cosmological data \cite{cosm_bimetric, recent_bimetric,sssmemh,bbems}.
However, the necessity to add a second metric-like field appears to be rather inelegant and is not well founded from the point of view of Riemannian 
geometry with the interaction potential between the two metrics introduced {\em ad hoc},  despite being motivated by non-linear generalizations of Fierz-Pauli massive gravity \cite{FP39} which do not suffer from Boulware-Deser ghost problem \cite{hr1, hr2}.  

Surprisingly, the hint of a geometric explanation might come from models used in particle physics. In a quest to explain the structure of the Standard
Model, a purely geometric interpretation of its content was proposed by Alain Connes using the tools of noncommutative geometry \cite{connes1,connes2,ks97}. Taken seriously,
it explains the existence of different fermions and gauge interactions as related to geometry of a finite type, related to a finite-dimensional algebra
$\mathbb{C} \oplus \mathbb{H} \oplus M_3(\mathbb{C})$, with the derivation of the action linked to a general principle of Euclidean spectral action,
which provides all terms, including the Yang-Mills-Higgs leading to the spontaneous symmetry breaking as well the pure gravity  Einstein-Hilbert action.

A simplified model of this type, which was the first considered \cite{CoLo} in the early days of the development of the theory, describes a product
of the smooth geometry (a four-dimensional manifold) with a two-point space. Such two-sheeted geometry, with a product structure is tractable in
noncommutative geometry leading to a simple Yang-Mills-Higgs toy model. However, from the point of view of gravity an interesting question is 
whether it is admissible to have different metrics on the two separate sheets of this geometry. This question is a challenge not only from the conceptual 
but also from the technical point of view, as it requires the computation of the spectral action in a much more general case than the product geometry.
In particular, the first question posed is whether the two metrics interact with each other.  A first step in this direction was done in \cite{sitarz2019}, 
where  a simple model of two Friedmann-Lema{\^i}tre-Robertson-Walker-type, flat geometries with identical lapse function was considered, resulting 
in the effective potential term linking the two geometries.

The present paper goes well beyond the restricted situation of the previous analysis, providing a full derivation of the potential linking the metric
and the equations of motion as well as the analysis of their stability. Though our model differs significantly from the typical bimetric theory (none
of the metrics can be thought of a background metric) the obtained potential is much similar to the bimetric case (though it is expressed as a rational
function and not a polynomial in the eigenvalues of the metrics ratio). Moreover, the symmetric coupling to the matter and radiation makes it
closer to the symmetric bimetric theory, where both metrics couple (in the same way) to matter and radiation. 

The paper is organized as follows: we present the assumptions of our model (the structure of the two-sheeted geometry) and the methods of
deriving the leading two terms of the spectral action using the pseudodifferential calculus and the Wodzicki residue. After computing the Euclidean
action functional for flat as well for curved geometries we perform the Wick rotation and obtain a set of nonlinear differential equations for the four 
functions that describe the model. In the rest we focus on the stability of the symmetric solutions, which are the standard Friedmann-Lema{\^i}tre-Robertson-Walker 
geometries for both sheets and analyse small perturbations for the typical cosmological solutions of dark-energy, matter and radiation-dominated
universes. In the last section we briefly discuss the possible physical consequences and argue why the model is physically viable.

\section{Almost commutative Friedmann-Lema{\^i}tre-Robertson-Walker models}

\subsection{Almost-commutative geometries}

The Gelfand-Naimark equivalence between topological spaces and commutative \mbox{$C^\ast$-algebras} was further enriched by A.~Connes in order to include noncommutative algebras and also to describe more than only the topology. In his formulation of noncommutative geometry \cite{connes85} the crucial role is played by a spectral triple which is a system $(\mathcal A,H,D)$ consisting of an unital $\ast$-algebra $\mathcal A$, Hilbert space $H$ and a Dirac type selfadjoint operator acting on $H$. Usually, more additional structure is assumed (e.g. the existence of grading-type operator $\gamma$, and an anti-unitary operator $J$, called real structure) and further compatibility conditions between all these elements. The canonical commutative example is $(C^{\infty}(M),L^2(M,S),D_M)$, where $M$ is a manifold equipped with a spin structure, $L^2(M,S)$ is the Hilbert space of square-integrable spinors, and $D_M=i\gamma^\mu(\partial_\mu+\omega_\mu)$ is the canonical Dirac operator expressed in the terms of the connection $\omega_\mu$ on the spinor bundle.

It turns out that crucial from the applications in particle physics point of view are triples with algebras being tensor products of the above one which some finite-dimensional matrix algebras $A_F$. The Hilbert space is the tensor product of $L^2(M,S)$ with some finite-dimensional Hilbert space $H_F$ on which $A_F$ is represented, and its dimension determines the number of fermionic degrees of freedom in the theory. Grading operators and real structures are also composed in an appropriate way in order to define analogous objects on the resulting triple. The Dirac operator, however, is not just the simple tensor product of $D_M$ and $D_F$, but has the following form:
\begin{equation}
D=D_M\otimes 1+\gamma_M\otimes D_F.
\end{equation}
The resulting triple forms the so-called almost-commutative geometry and have been the backbone of multiple models applied to the physics of elementary particles
(see \cite{WvS,DKL19}). The starting point to consider physical models based on spectral triples  is the spectral action. Its bosonic part is given by
\begin{equation}
\mathcal{S}(D)=\mathrm{Tr} f\left(\frac{D}{\Lambda}\right),
\end{equation}
where $\Lambda$ is some cut-off parameter and $f$ is some smooth approximation of the characteristic function of the interval $[0,1]$.  In the case of particle physics 
models it reproduces the bosonic part of the Lagrangian of such theories minimally coupled to gravity, together with the standard Hilbert-Einstein action for the metric.

\subsection{The classical geometry}
We consider geometries described by the generalized Friedmann-Lema{\^i}tre-Robertson-Walker metric, 
\begin{equation}
ds^2=b(t)^2 dt^2+a(t)^2\left(d\chi^2 +S_k^2(\chi)\left(d\theta^2 +\sin^2(\theta)d\phi^2\right)\right),
\end{equation}
where
\begin{equation}
S_k(\chi)=\begin{cases}\sin(\chi), \qquad &k=1,\\
\chi,\qquad &k=0,\\
\sinh(\chi), \qquad &k=-1\end{cases}
\end{equation}
and $a(t),b(t)$ are positive (sufficiently smooth) functions. 

The orthogonal coframe $\{\theta^a\}$ for $ds^2$ is defined so that $ds^2=\theta^a\theta^a$. It allows us to immediately compute 
the spin connection $\omega$ which is determined by $d\theta^a=\omega^{ab}\wedge \theta^b$. Then, the Dirac operator is, in a local 
coordinates, given by 
\begin{equation}
\label{dirac}
D=\gamma^a dx^\mu(\theta_a)\frac{\partial}{\partial x^\mu} +\frac{1}{4}\gamma^c \omega_{cab}\gamma^a\gamma^b,
\end{equation}
where $\gamma^a$'s are gamma matrices chosen to be antihermitian and so that $\gamma^{a}\gamma^{b}+\gamma^{b}\gamma^{a}=-2\delta^{ab}I$. 

Instead of the original Dirac operator we can equivalently analyse the operator, which
is conformally rescaled, $D_h=h^{-1}Dh$, with the scale factor $h(t)=a(t)^{-3/2}b(t)^{-1/2}$.
This assures that we can work with the Hilbert space of spinors, where the scalar product 
does not depend on $a(t)$ and $b(t)$.

\subsection{The two-sheet almost commutative model}
We consider a generalised almost-commutative geometry, which is described by a product-like spectral triple 
of the spectral triple over the manifold with the Friedmann-Lema{\^i}tre-Robertson-Walker metric and the triple over two points. However, instead of 
the usual product Dirac operator, we take a more general one,
\begin{equation}
\mathcal D=\begin{pmatrix}
D_1 & \gamma \Phi \\
\gamma \Phi^\ast & D_2
\end{pmatrix},
\end{equation}
where $D_1,D_2$ are both of the form \eqref{dirac}, yet with possibly different scaling functions $a$ and $b$, and $\Phi$ being 
a priori a field (which can be later restricted to be constant). 

The choice of the full Dirac operator with the $\gamma$ in the off-diagonal part is motivated by the fact that in the case of $D_1=D_2$ it yields 
a usual almost-commutative  product geometry. Note that, in principle one can study generalized objects with arbitrary order-zero operators 
on the off-diagonal of ${\mathcal{D}}$, so the only thing we require of $\gamma$ is that it anticommutes with $\gamma^a$ matrices 
 and  is not necessarily the chirality grading operator of the Euclidean spin geometry of the manifold $M$.  In order to have the full Dirac
operator hermitian we must require that $\gamma$ is hermitian and, consequently, we have to normalize $\gamma^2=1$.  However, we
shall relax this assumption and consider also models with $\gamma^2=-1$. This allows for much more flexibility, in particular, for the models 
that are derived from higher-dimensional Kaluza-Klein type geometries and would lead to some more realistic effective physical situations. 
One of the interesting possibilities is that when passing to the  Lorentzian   signature for the manifold $M$ we can as well choose the  Lorentzian  
signature for the discrete degrees of freedom. This possibility has been discussed for finite geometries, albeit in a different context of the Standard 
Model in \cite{BS18}, in the natural language  of  Krein spaces. What is important for our consideration is that the only difference will be that 
the operator ${\mathcal{D}}$ will be only Krein  self-adjoint,  meaning that $\gamma$ will be antiselfadjoint and $\gamma^2=-1$.  
To accommodate for both possibilities in the discrete degrees of freedom we do not fix $\gamma^2$ and we allow that (after normalization) 
$\gamma^2 = \kappa = \pm 1$. 

To simplify the presentation in the paper we introduce the following matrices:
\begin{equation}
B(t)=\begin{pmatrix}
\frac{1}{b_1(t)} & \\
& \frac{1}{b_2(t)}
\end{pmatrix}, \hspace{10pt}
A(t)=\begin{pmatrix}
\frac{1}{a_1(t)} & \\
& \frac{1}{a_2(t)}
\end{pmatrix},
\hspace{10pt}
F(t,x)=
\begin{pmatrix}
 & \Phi(t,x) \\
 \Phi(t,x)^\ast &
\end{pmatrix}.
\end{equation}

\subsection{The spectral action}
For the geometry described by a given Dirac operator $D$ the main object of interest is the Laplace-type 
operator $D^2$, which is a second-order differential operator acting on the sections of the doubled spinor bundle. 
Its symbol $\sigma_{D^2}(x,\xi)$ consists of three parts ${\mathfrak a}_0+ {\mathfrak a}_1+  {\mathfrak a}_2$, each 
of ${\mathfrak a}_k(x,\xi)$ being homogeneous of degree $k$ in $\xi$'s. Then we compute the symbol of its inverse,
\begin{equation}
\sigma_{D^{-2}}(\xi)={\mathfrak b}_0+{\mathfrak b}_1+{\mathfrak b}_2+...,
\end{equation}
where ${\mathfrak b}_k(x,\xi)$ is homogeneous of order $-\!2\!-\!k$ in $\xi$
(we briefly review the mathematical details of how the computations of the symbols are performed in the 
Appendix~\ref{sec:symbols}) and use it to compute the first two terms of the spectral action for the considered model. 

It can be expressed in terms of Wodzicki residua \cite{kastler, kw} as,
\begin{equation}
\mathcal{S}(D)=\Lambda^4 \, \mathrm{Wres}(D^{-4})+c\Lambda^2 \, \mathrm{Wres}(D^{-2})=\int_M\int_{\|\xi\|=1} 
\left(\Lambda^4\, \mathrm{Tr}\, \mathrm{Tr}_{Cl} \, {\mathfrak b}_0^2+c \Lambda^2 \, \mathrm{Tr}\, \mathrm{Tr}_{Cl} {\mathfrak b}_2 \right),
\end{equation}
where $\mathrm{Tr}_{Cl}$ denotes the trace performed over the Clifford algebra and $\mathrm{Tr}$ is the trace over the 
matrices $M_2(\CC)$ that are used in the mild noncommutativity introduced in the model. 

\subsection{Flat geometries.}\label{sec:fg}
Although the topology of the flat case in physics is not exactly toroidal, from the point of view  of local behaviour it is 
identical to such, which was already analysed for $b=1$ in \cite{sitarz2019}. In this section we generalize those results 
to the case with arbitrary function $b(t)$, so we consider here toroidal Friedmann-Lema{\^i}tre-Robertson-Walker geometries described 
by the following metric in the coordinate system $(t,x)=(t,x^1,x^2,x^3)$:
\begin{equation}
ds^2=b(t)^2dt^2+a(t)^2\left((dx^1)^2+(dx^2)^2+(dx^3)^2\right).
\end{equation}
Hence an orthogonal frame for $ds^2$ is of the form,
\begin{equation}
\theta^0=b(t)dt, \qquad  \theta^1=a(t)dx^1,\qquad \theta^2=a(t)dx^2, \qquad \theta^3=a(t)dx^3,
\end{equation}
while the matrix of connection $1$-forms is,
\begin{equation}
\omega=\frac{1}{a(t)b(t)}\begin{pmatrix}
0 & -(\partial_ta)\theta^1 & -(\partial_t a)\theta^2 & -(\partial_t a)\theta^3 \\
(\partial_t a)\theta^1 & 0& 0& 0 \\
(\partial_t a)\theta^2 & 0& 0& 0 \\
(\partial_t a)\theta^3 & 0& 0& 0 \\
\end{pmatrix}.
\end{equation}
As a result, the (single) Dirac operator takes the following form,
\begin{equation}
D=\frac{1}{b(t)}\gamma^0\left(\partial_t + \frac{3\partial_ta}{2a(t)}\right)+\frac{1}{a(t)}\gamma^j\partial_j,
\end{equation}
and after the conformal rescaling $h(t)=a(t)^{-3/2}b(t)^{-1/2}$ we get
\begin{equation}
D_h=\frac{1}{b(t)}\gamma^0\left(\partial_t-\frac{\partial_t b}{2b(t)}\right)+\frac{1}{a(t)}\left(\gamma^1\partial_1+\gamma^2\partial_2+\gamma^3\partial_3\right),
\end{equation}
so that the full Dirac operator acting on the doubled Hilbert space of spinors is,
\begin{equation}
\mathcal D=\gamma^0\left(B(t)\partial_t-\partial_t B\right) + A(t) \gamma^j\partial_j +\gamma F(t,x).
\end{equation}
The resulting Laplace-type operator in this model is of the following form:
\begin{equation}
\begin{aligned}
\mathcal D^2=&-B^2\partial_t^2-A^2\partial^2+B(\partial_t A)\gamma^0\gamma^k\partial_k+[F,A]\gamma\gamma^k\partial_k \\
&+ [F,B]\gamma\gamma^0\partial_t+ \kappa F^2 +\gamma^0\gamma B(\partial_t F)+\gamma^j\gamma A(\partial_j F)  \\
&+ \gamma^0\gamma[F,\partial_t B]+B(\partial_t^2 B) + B(\partial_t B)\partial_t -(\partial_t B)^2.
\end{aligned}
\end{equation}
The symbol $\sigma(\mathcal D^2)={\mathfrak a}_0+{\mathfrak a}_1+{\mathfrak a}_2$ is given by
\begin{equation}
\begin{aligned}
&{\mathfrak a}_2= B^2\xi_0^2 +A^2\xi^2,\\
&{\mathfrak a}_1= i\left[B(\partial_t A)\gamma^0\gamma^k\xi_k+B(\partial_t B)\xi_0
  +[F,A]\gamma\gamma^k\xi_k+[F,B]\gamma\gamma^0\xi_0\right],\\
&{\mathfrak a}_0= \kappa F^2-\gamma\gamma^0\left(B(\partial_tF)+[F,\partial_t B]\right)
  -\gamma\gamma^jA(\partial_jF)+B(\partial_t^2 B)-(\partial_t B)^2,
\end{aligned}
\end{equation}
where we denoted by $\xi^2=\xi_1^2+\xi_2^2+\xi_3^2$. Now, computing the symbol of 
$\mathcal{D}^{-2}$ using the prescription presented in Appendix \ref{sec:symbols}, we obtain ${\mathfrak b}_0(\mathcal{D})$ and ${\mathfrak b}_2(\mathcal{D})$. Then, taking the trace over the Clifford
algebra and the matrices $M_2(\mathbb{C})$, and integrating over the cosphere bundle $|\xi|^2=1$, we compute the Wodzicki residue that gives us the Euclidean 
spectral action of the considered model. The final result is,
\begin{equation}
\begin{aligned}
\mathcal{S}(\mathcal D)&\sim\int dt \left\{\Lambda^4 (a_1^3b_1+a_2^3b_2) - \frac{c\Lambda^2}{12} 
\left( a_1^3 b_1 R(a_1,b_1) + a_2^3 b_2 R(a_2, b_2) \right) \right.  \\
& \left. +c \kappa \Lambda^2|\Phi|^2b_1b_2\frac{(a_1-a_2)^2}{(a_1b_2+a_2b_1)^2}\left[a_1^2(2a_2b_1+a_1b_2)+a_2^2(2a_1b_2+a_2b_1)\right] \right.\\
& \left. + c \kappa \Lambda^2|\Phi|^2\frac{(b_1-b_2)^2}{(a_2b_1+a_1b_2)^2}a_1^2a_2^2(a_1b_1+a_2b_2)
-c \kappa \Lambda^2|\Phi|^2(a_1^3b_1+a_2^3b_2),\right. 
\end{aligned}
\label{S-k0}
\end{equation}
where the scalar curvature for the flat spatial geometry is,
\begin{equation}
R(a,b) =6 \left( \frac{\partial_t a \, \partial_t b}{a b^3} - \frac{ (\partial_t a)^2}{a^2 b^2} - \frac{\partial_t^2 a}{a b^2} \right).
\end{equation}
%%%%%%%%%%%%%%%%%%%%%%%%%%%%%%%%%%%%%%%%%%%%%%%%%%%%%%%%%%%%%%%%%%%%
\subsection{The non-flat case}
In this subsection we concentrate on the case with positive ($k\!=\!1$) curvatures, with the negative ($k\!=\!-1$) case that can be 
treated in a similar manner. Although the effective Lagrangian and the equations of motion are local and hence the dynamical terms 
are expected to be unchanged, we derive them explicitly using appropriate coordinates.  For the case of $k=1$ we use the spherical 
coordinates $(t,\chi,\theta,\phi)$, so that  the metric is then described by:
\begin{equation}
ds^2=b(t)^2 dt^2+a(t)^2\left(d\chi^2 +\sin^2(\chi)\left(d\theta^2 +\sin^2(\theta)d\phi^2\right)\right).
\end{equation}

The orthogonal frame is given by 
\begin{equation}	
\theta^0=b(t)\,dt, \quad \theta^1=a(t)\,d\chi, \quad
\theta^2=a(t)\sin\chi \,d\theta, \quad
\theta^3=a(t)\sin\chi \sin\theta \,d\phi,
\end{equation}
hence
\begin{equation}
\begin{aligned}
& d\theta^0=0, \qquad  d\theta^1=\frac{\partial_t a}{ab}\theta^0\wedge \theta^1, \qquad d\theta^2=\frac{\partial_t a}{ab}\theta^0\wedge\theta^2+\frac{\cot\chi}{a}\theta^1\wedge\theta^2 \\
& d\theta^3=\frac{\partial_t a}{ab}\theta^0\wedge\theta^3+\frac{\cot\chi}{a}\theta^1\wedge\theta^3+\frac{\cot\theta}{a \sin\chi}\theta^2\wedge \theta^3.
\end{aligned}
\end{equation}

Therefore the only nonvanishing components for the spin connection $\omega$ are \cite{cc2012, fgk2014}:
\begin{equation}
\omega_{101}=\omega_{202}=\omega_{303}=\frac{\partial_t a}{ab},  \quad \omega_{212}=\omega_{313}=\frac{\cot\chi}{a},\qquad\omega_{323}=\frac{\cot\theta}{a \sin\chi}.
\end{equation}
Now, for the Dirac operator we get explicitly
\begin{equation}
D=\gamma^0\frac{1}{b}\left(\frac{\partial}{\partial t}+\frac{3}{2}\frac{\partial_t a}{a}\right)+ \frac{1}{a}D_3,
\end{equation}
where in this case
\begin{equation}
D_3=\gamma^1\frac{\partial}{\partial \chi}+\gamma^2 \csc\chi \frac{\partial}{\partial \theta}+\gamma^3\csc\chi\,\csc\theta\frac{\partial}{\partial\phi}+\gamma^1\cot\chi+\frac{1}{2}\gamma^2\cot\theta\,\csc\chi.
\end{equation}
After the conformal rescaling by using $h(t)=a(t)^{-3/2}b(t)^{-1/2}$ we end up with the following Dirac operator
\begin{equation}
D_{h}=\frac{1}{b}\gamma^0\left(\frac{\partial}{\partial t}-\frac{\partial_t b}{b}\right)+\frac{1}{a}D_3,
\end{equation}
Therefore, for the doubled model that we are considering, the Dirac operator is
\begin{equation}
\mathcal{D}=\gamma^0\left(B(t)\partial_t-\partial_t B\right) + A(t)D_3+\gamma F(t,x).
\end{equation}
As a result we have
\begin{equation}
\begin{aligned}
{\mathcal D}^2=&-B^2\partial_t^2+A^2 D_3^2+ B(\partial_t A)\gamma^0D_3 + [F,A]\gamma D_3 \\
&+[F,B]\gamma\gamma^0\partial_t + \kappa F^2  +\gamma^0\gamma B(\partial_t F)-\gamma A(D_3F) \\
&+\gamma^0\gamma[F,\partial_t B] + B(\partial_t^2 B)+B(\partial_t B)\partial_t -(\partial_t B)^2.
\end{aligned}
\end{equation}
In order to compute its symbol $\sigma_{\mathcal D^2}={\mathfrak a}_2+{\mathfrak a}_1+{\mathfrak a}_0$ 
we first notice that the symbol of $D_3^2$ is given by:
\begin{equation}
\begin{aligned}
{\mathfrak a}_2(D_3^2) \,=\,&\xi_\chi^2+\csc^2\chi\, \xi_\theta^2+\csc^2\chi\, \csc^2\theta \, \xi_\phi^2, \\
{\mathfrak a}_1(D_3^2)\, =\,&-i\left(2\cot\chi\, \xi_\chi+\cot\theta\, \csc^2\chi \, \xi_\theta +\gamma^1\gamma^2\cot\chi\, \csc\chi\, \xi_\theta + \right. \\
&\qquad +\left. \gamma^1\gamma^3 \csc\theta\, \cot\chi \, \csc\chi \, \xi_\phi +\gamma^2\gamma^3 \cot\theta\, \csc\theta\, \csc^2\chi \, \xi_\phi \right), \\
{\mathfrak a}_0(D_3^2) \,=\, &-\frac{1}{2}\gamma^1\gamma^2\cot\theta\, \cot\chi\, \csc\chi + \csc^2\chi-\cot^2\chi 
\\ & \qquad +\frac{1}{2}\csc^2\theta\, \csc^2\chi - \frac{1}{4}\cot^2\theta\, \csc^2\chi.
\end{aligned}
\end{equation}
As a result, for the operator $\mathcal D^2$, we have
\begin{equation*}
\begin{aligned}
{\mathfrak a}_2 \,=\, &B^2\xi_0^2+A^2\xi_\chi^2+\csc^2\chi\, A^2\xi_\theta^2+\csc^2\chi\, \csc^2\theta \, A^2\xi_\phi^2, \\
{\mathfrak a}_1 \,=\, &-i\left\{2\cot\chi\, A^2\xi_\chi+\cot\theta\,\csc^2\chi\, A^2\xi_\theta-B(\partial_t B)\xi_0+\right.  \\
& \quad + \gamma^1\gamma^2 A^2 \cot\chi \, \csc\chi\, \xi_\theta + \gamma^1\gamma^3A^2\csc\theta\, \cot\chi\,\csc\chi\, \xi_\phi \\ 
& \quad +\gamma^2\gamma^3 A^2\cot\theta\, \csc\theta\, \csc^2\chi \, \xi_\phi - B(\partial_t A) \gamma^0\gamma^1\xi_\chi  \\
& \quad -B(\partial_t A)\gamma^0\gamma^2\csc\chi\, \xi_\theta -B(\partial_t A)\gamma^0\gamma^3 \csc\chi\, \csc\theta\, \xi_\phi - \\
& \quad - [F,A]\gamma\gamma^1\xi_\chi-[F,A]\gamma\gamma^2\csc\chi\, \xi_\theta -[F,A]\gamma\gamma^3 \csc\chi\, \csc\theta\, \xi_\phi \\
& \quad \left. -[F,B]\gamma\gamma^0\xi_0\right\},
\end{aligned}
\end{equation*}
\begin{equation*}
\begin{aligned}
%%%%%%%%%%%%%%%%%%%%%%%%%%%%%%%%%%
{\mathfrak a}_0=&A^2\left(\csc^2\chi-\cot^2\chi +\frac{1}{2}\csc^2\theta\, \csc^2\chi - \frac{1}{4}\cot^2\theta\, \csc^2\chi\right) \\
& \quad + \kappa F^2+B(\partial_t^2 B)-(\partial_t B)^2 -  \frac{1}{2} \gamma^1\gamma^2\cot\theta\, \cot\chi\, \csc\chi \\
& \quad + B(\partial_t A)\gamma^0\gamma^1 \cot\chi +\frac{1}{2}B(\partial_t A)\gamma^0\gamma^2\cot\theta\, \csc\chi\\
& \quad +[F,A]\gamma\gamma^1\cot\chi +\frac{1}{2}[F,A]\gamma\gamma^2\cot\theta\, \csc\chi +\gamma^0\gamma [F,\partial_t F] \\
& \quad - \gamma\gamma^0B(\partial_t F)-\gamma\gamma^1 A \left(\partial_\chi F +F \cot\chi \right)-\gamma\gamma^2A \csc\chi\, \\ 
& \quad + \left(\partial_\theta F  +\frac{F}{2}\cot\theta\right) -\gamma\gamma^3 A \csc\chi\, \csc\theta\, \partial_\phi F.
\end{aligned}
\end{equation*}
Using the prescription presented in Appendix \ref{sec:symbols}, we first compute the symbols $\sigma_{\mathcal D^{-2}}={\mathfrak b}_0+{\mathfrak b}_1+{\mathfrak b}_2+...$, 
then we proceed, in an exactly similar manner as in the case of the toroidal geometry, to compute the spectral action. The result is,
\begin{equation}
\begin{split}
\mathcal{S}(\mathcal D)&\sim\int dt \left\{\left(\Lambda^4 -c \kappa \Lambda^2|\Phi|^2\right)(a_1^3b_1+a_2^3b_2) - \frac{c\Lambda^2}{12}\left(a_1^3 b_1 R(a_1,b_1)+a_2^3b_2 R(a_2,b_2)\right)\right.  \\
& \left. +c \kappa \Lambda^2|\Phi|^2b_1b_2\frac{(a_1-a_2)^2}{(a_1b_2+a_2b_1)^2}\left[a_1^2(2a_2b_1+a_1b_2)+a_2^2(2a_1b_2+a_2b_1)\right]+\right.\\
& \left. + c \kappa \Lambda^2|\Phi|^2\frac{(b_1-b_2)^2}{(a_2b_1+a_1b_2)^2}a_1^2a_2^2(a_1b_1+a_2b_2)
\right\},
\end{split}
\label{S-k1}
\end{equation}
where now $R(a,b)$ denotes the scalar of curvature for spherical spatial geometries, 
\begin{equation}
R(a,b) = 6 \left( \frac{ \partial_t a \, \partial_t b}{a b^3}-\frac{ (\partial_t  a)^2}{a^2 b^2}-\frac{ \partial_t^2 a}{a b^2}+\frac{1}{a^2} \right).
\end{equation}

Note that the action functional differs \eqref{S-k1} from \eqref{S-k0} only through the last term that arises from the scalar curvature 
of the spherical spatial geometry, where a relevant term depending on $k=1$ is added. The above result can be generalized to the negative 
curvature case (we omit straightforward  but tedious parametrization and computation of symbols). In fact, taking $R(a,b)$ as $R(a,b,k)$, 
depending on the space curvature $k$, we have a general action functional for all geometries in the doubled spacetime Friedmann-Lema{\^i}tre-Robertson-Walker models.

\subsection{The interactions of the metrics}\label{sG}
Before we pass to the equations of motions an their stability, let us briefly compare the effective potential describing the
interaction between the two metrics to the bimetric gravity models\cite{hr1, akms13, sssmemh}. 
Certainly, apart from the fact that we have an action for
two metrics, there is a much deeper symmetry between the two, since neither plays a role of a ,,background'' metric. In fact,
the usual solution in the case of vanishing $\alpha$ gives both metric  totally independent of each other. Introducing the variables,
$$ x = \frac{b_1}{b_2}, \qquad y= \frac{a_1}{a_2}, $$
which depend only on the entries of the matrix $X^a_{\ c}=g_2^{ab} {g_1}_{bc}$, we can express the interactions between the metrics as proportional to:

\begin{equation}
\label{bimetric_eq}
\mathbb{V}\left(\sqrt{g_2^{-1}g_1}\right)\sqrt{g_2} =\mathbb{V}\left(\sqrt{g_1^{-1}g_2}\right)\sqrt{g_1},
\end{equation}
where the function $\mathbb{V}\left(\sqrt{g_2^{-1}g_1}\right)$ is of the form:
$$
\frac{x^2 + 2xy -2x^2y +y^2-6xy^2 +4x^2y^2+4xy^3-6x^2y^3 +x^3y^3 -2xy^4+2x^2y^4+xy^5}{(x+y)^2}
$$ 
which can be efficiently expressed as a rational function of the symmetric polynomials in $\sqrt{X}$.

We stress that the resulting model possesses features that are characteristic to bimetric gravity models: the potential $\mathbb{V}$ depends on the metrics only 
through $\sqrt{X}$ and satisfies \eqref{bimetric_eq}. On the other hand, in the usual bimetric models such potential is a polynomial in eigenvalues of $\sqrt{X}$ rather 
than a rational function. It was proposed in \cite{ab2019} that the construction presented here might results in the derivation of bimetric theories out of the 
geometric data. The above result suggest that indeed this class of models resembles some characteristics of bimetric gravity models, but is a different one. 
We postpone for the future research the detailed analysis of these differences and their cosmological implications. 

\subsection{The equations of motion}

The action functional \eqref{S-k1} depends on the field $B$ only via $b_1$ and $b_2$ but not their derivatives. As a result, $b_1$ and $b_2$ are not dynamical and its Euler-Lagrange equations give rise to the constraints of the model. Moreover, due to the reparametrization invariance we can fix one of these functions 
or relate them with each other. 

Furthermore, the action functional was derived for the Euclidean model and to pass to physical situation we need to perform 
Wick rotation,  as described in the \cite{sitarz2019}. In our case, this will affect only the square of the time derivative of the scaling 
factors $a_i(t)$, which will change signs. Consequently, the action and the equation of motion for the rest of this paper are in the Lorentzian signature 
of the metric $(-,+,+,+)$. Let us remind that the discrete degrees of freedom of the geometry might be Riemannian or pseudo-Riemannian, which
results in the appropriate choice of the sign $\kappa$.

After integration by parts and omitting the boundary terms that are full derivatives in $t$, we obtain the following 
action for the pure gravity Friedmann-Lema{\^i}tre-Robertson-Walker doubled geometries for the Lorentzian signature and arbitrary spatial 
curvature $k$,
\begin{equation}
\label{S-k1L}
\begin{aligned}
\mathcal{S}_k(\mathcal D) = \left(\frac{c\Lambda^2}{12}\right)  &
\left\{  \int dt \left( \Lambda_e (a_1^3b_1+a_2^3b_2)   - 6k  \left( a_1 b_1 + a_2 b_2 \right) \right. \right.   \\
& +6 \left(\frac{a_1}{b_1}(\partial_t a_1)^2+\frac{a_2}{b_2}(\partial_t a_2)^2\right) \\
&  +\alpha \,b_1 b_2\frac{(a_1-a_2)^2}{(a_1b_2+a_2b_1)^2}\left[ a_1^2(2a_2 b_1+a_1 b_2)+a_2^2 (2 a_1 b_2+a_2 b_1) \right] \\
& \left. \left. + \alpha \,\frac{(b_1-b_2)^2}{(a_2 b_1+a_1 b_2)^2} a_1^2 a_2^2(a_1  b_1+a_2 b_2) \right) \right\},
\end{aligned}
\end{equation}
where we have factored out the overall constant so that the dynamical term appears only with a numerical
factor, denoted the effective cosmological constant by $\Lambda_e$ and introduced the effective coupling
between the two metrics by $\alpha$: 
$$ \Lambda_e = \frac{12}{c} \left( \Lambda^2 - c \kappa |\Phi|^2 \right),
\qquad \alpha = 12 |\Phi|^2 \kappa. $$ 
Unlike the bare cut-off parameter $\Lambda$, here the effective cosmological constant can vanish or be negative for a particular model.  
We shall use the above convention with $\Lambda_e$ and $\alpha$ throughout the rest of the paper.

The four Euler-Lagrange equations take the following form,
\begin{equation}
\label{EL01}
\Lambda_e= 6 H_{b,i}^2 + 6 \frac{k}{a_i^2}
- \frac{\alpha}{a_i}  V(a_i,a_{i'}, b_i, b_{i'}), 
\end{equation}
with 
$$ 
V(a_1,a_2,b_1,b_2) = a_1+\frac{8a_1 a_2 (a_1^2 - a_2^2) b_2^3}{(a_2 b_1+a_1 b_2)^3}
+\frac{2 a_2 (a_2^2+2 a_1 a_2-5 a_1^2) b_2^2}{(a_2 b_1+a_1 b_2)^2}, $$
and
\begin{equation}
\label{EL03}
12 \frac{\partial_t^2{a}_i}{a_i b_i^2}+6H_{b,i}^2
-3\Lambda_e + 6 \frac{k}{a_i^2} -12\frac{(\partial_t {a}_i)(\partial_t{b}_i)}{a_i b_i ^3} -
\alpha W(a_i,a_{i'}, b_i, b_{i'}) = 0,
\end{equation}
with 
$$
\begin{aligned}
W(a_1,a_2,b_1,b_2) = & \,
3 - 2 \frac{a_2 b_2 (a_2^2 - 4 a_1 a_2 + 9 a_1^2)}{a_1^2(a_2b_1+a_1b_2)}
+ 2 \frac{a_2 b_2^2 (11 a_1^2 - 2 a_1 a_2 -3 a_2^2)}{a_1(a_2b_1+a_1b_2)^2} \\
& - 8 \frac{a_2 b_2^3 (a_1^2 -  a_2^2)}{(a_2b_1+a_1b_2)^3}.
\end{aligned}
$$
In the above equations we use the convention that $(i,i') = \{ (1,2), (2,1) \},$ and
$H_{b,j}=\frac{\partial_t{a}_j}{a_jb_j}$ are  the generalized Hubble parameters.
%%%%%%%%%%%%%%%%%%%%%%%%%%%%%%%%%%%%%%%%%%%%%%%%%%%
Before we analyse the inclusion of matter fields and possible solutions, let us observe that
in the flat $k\! = \! 0$ case, inserting $\Lambda_e$ from first two equations in last two, 
one obtains
\begin{equation}
\begin{aligned}
\frac{6}{b_1}\partial_t{H}_{b,1}+\alpha a_2b_2\frac{a_2b_1-a_1b_2}{a_1^2} L(a_1,a_2,b_1,b_2)=0, \\
\frac{6}{b_2}\partial_t{H}_{b,2}+\alpha a_1b_1\frac{a_1b_2-a_2b_1}{a_2^2} L(a_1,a_2,b_1,b_2)=0,
\end{aligned}
\end{equation}
with some rational function $L(a_1,a_2,b_1,b_2)$, so in particular, whenever $a_1b_2=a_2b_1$ 
both $H_{b,1}$ and $H_{b,2}$ must be constant. 
%%%%%%%%%%%%%%%%%%%%%%%%%%%%%%%%%%%%%%%%%%%%%%%%%%%
\section{Interaction with matter fields and radiation}\label{s3}

The equation of motion derived in the previous section describe the empty universe in the doubled model. Here, 
we can ask how they are modified by the presence of the matter fields. The crucial point is to see how the effective
matter and radiation action depend on the components of the metrics described in terms of fields $a_1,a_2$,
$b_1,b_2$. The main difficulty is the passage from the microscopic action for spinor and gauge fields to the
effective averaged energy-momentum tensor in the Einstein equations. 

The microscopic action for the spinor fields in the doubled universe will be the usual fermionic action 
$ \overline{\Psi} {\mathcal D} \Psi $. Since both components of the spinor couple to the respective
Dirac operators $D_1$ and $D_2$ on each of the single sheets separately, and the $\Psi$ field is, by assumption,
independent of the metric fields, we conclude that the resulting action will be split into separate actions that
do not mix the metric components on each of the single universes.

Similar argument can be used for the radiation energy-momentum tensor that originates from the gauge fields
over the considered model. As the model has two $U(1)$ symmetries there are two gauge fields that couple to the Higgs
field. A linear combination of them will become a massive one, due to spontaneous symmetry breaking of the 
Higgs field, whereas another linear combination will correspond to the massless photons. Again, the effective
Yang-Mills action for the photon field will not mix the metric components over the two sheets and therefore we shall
have independent tensor-energy components for each equation. 

These heuristic arguments suggest that the effective equations of motion are modified by the respective components
of the overall energy-momentum tensor $T^0_{\ 0}$ and $T^1_{\ 1}$, which depend separately on $a_1,b_1$ and $a_2, b_2$, 

\begin{equation}
\begin{aligned}
\label{ELkk}
& 6 H_{b,i}^2 +\frac{6k}{a_i^2} -\Lambda_e - \frac{\alpha}{a_i} V(a_i,a_{i'},b_i,b_{i'}) = - 2 T^0_{\ 0}(a_i,b_i), \\
&  12 \frac{\partial_t^2{a}_i}{a_ib_i^2} + 6H_{b,i}^2 - 3\Lambda_e + \frac{6k}{a_i^2} 
- 12\frac{\partial_t{a}_i\, \partial_t{b}_i}{a_i b_i^3} - \alpha W(a_i,a_{i'},b_i,b_{i'}) = -6 T^1_{\ 1}(a_i,b_i),
\end{aligned}
\end{equation}
for $(i,i')=\{ (1,2), (2,1)\}$.

As in the conventional cosmology we consider the model of the perfect fluid, i.e. the stress-energy tensor is taken to 
be of the form
\begin{equation}
T^g_{\mu\nu}=(\rho+P)u_\mu u_\nu +P g_{\mu\nu},
\end{equation} 
where $P$ is refered to as pressure, while $\rho$ is called energy density. For the generalized Friedmann-Lema{\^i}tre-Robertson-Walker metric, the vector $u^{\mu}$ is $\left(\frac{1}{b(t)},0,0,0\right)$, so that $u_\mu u^\mu=-1$. As a result, $T^0_{\ 0}=-\rho$ and $T^1_{\ 1}=P$. 

Furthermore, the continuity equation $\nabla_\mu T^{\mu\nu}=0$ reduces to the standard one:
\begin{equation}
\frac{\partial \rho}{\partial t}+3(\rho+P)\frac{\partial_t{a}}{a}=0.
\end{equation}
We assume that the thermodynamics of the matter content is characterized by the following equation of state:
\begin{equation}
P(t)=w \rho(t).
\end{equation}
From the continuity equation we immediately infer that then 
\begin{equation}
\rho(t)=\eta\, a(t)^{-3(1+w)},
\end{equation}
where $\eta$ is the proportionality constant, exactly as in the standard cosmology.

The resulting Einstein equations for the double-sheeted universe are of the following form:

\begin{equation}
\begin{aligned}
\label{EL1kk}
& 6 H_{b,i}^2 +\frac{6k}{a_i^2} -\Lambda_e - \frac{\alpha}{a_i} V(a_i,a_{i'},b_i,b_{i'}) = \frac{2\eta}{a_i^{3(1+w)}} \\
&   12 \frac{\partial_t^2{a}_i}{a_ib_i^2} + 6H_{b,i}^2 - 3\Lambda_e + \frac{6k}{a_i^2} 
- 12\frac{\partial_t{a}_i\, \partial_t{b}_i}{a_i b_i^3} - \alpha W(a_i,a_{i'},b_i,b_{i'}) = -\frac{6w\eta}{a_i^{3(1+w)}},
\end{aligned}
\end{equation}
for $(i,i')=\{ (1,2), (2,1)\}$.

We stress that the above model is a straightforward generalization of the classical one for the doubled theory, with the
only difference that we allow two different scaling factors and the interaction between them derived from the spectral
action.  Indeed, for $a_1=a_2\equiv a$, $b_1=b_2=1$, or $\alpha=0$, equations of motion reduces to the usual 
Friedmann equations yielding the well-known solutions. 

In what follows we shall aim to analyse the possibility of small perturbations of the classical solutions of Friedmann-Lema{\^i}tre-Robertson-Walker models,
trying to answer the question whether the double-sheeted universe is stable. However, before we start the computations
to see when the small perturbations of the classical solution are possible, let us observe that thanks to the reparametrisation
invariance of the time variable in the equations  \eqref{EL01}, \eqref{EL03}, we can decide to fix either $b_1$ or $b_2$ or relate 
them with each other. There are, in principle, many choices of the possible parametrizations and we choose a particular
one, which is motivated by the existing symmetry  with respect to the  exchange between left and right modes 
in the geometric setup we consider. We shall set  $b_1(t) + b_2(t)= 2$, therefore, effectively, one can introduce 
a new function,
$$ b_1(t) = 1+ b(t), \qquad b_2(t) = 1 - b(t), $$
and derive the equations of motion for $a_1(t), a_2(t)$ and $b(t)$. Taken an appropriate linear
combination of the derivatives of \eqref{EL01} and the equations \eqref{EL03} we shall obtain
three nonlinear first order differential equations for these functions. 

Despite the fact that a full analysis of these equations is complicated and can be done possibly only 
numerically, we can obtain some significant results. 

We shall finish this section by a remark that one cannot a'priori assume that both lapse functions are identically $1$. Indeed,
we shall see that such solutions (in the linearised regime) are not possible. Moreover, a far more general argument holds also
for the full equations in the case of the empty universe. Then, there are no solutions with $b(t)=0$ apart from $a_1(t)=a_2(t)$.
The argument is quite simple and relies on algebraic manipulation of the equations \eqref{EL01}. Indeed, assuming 
$b_1(t)=1=b_2(t)$ and subtracting the equations we obtain the following relation between $a_1$ and $a_2$,
$$ 2 \alpha \frac{a_1-a_2}{(a_1+a_2)^2} \left( a_1^2 + 4 a_1 a_2 +a_2^2 \right)\left(a_2^2 \, \partial_t {a_1}+  a_1^2\, \partial_t {a_2}\right)  = 0, $$
which is true only if $a_1=a_2$ (since both $a_1,a_2$ are positive functions) or $\frac{1}{a_1} + \frac{1}{a_2} = \hbox{const}$. 
The latter condition can be solved, and when used in either of the first two equations \eqref{EL01} it leads to the constant solutions for 
$a_1$ and $a_2$. Therefore, the functional based on the action \eqref{S-k1L} has extremal points only if the time scaling factor differs 
for the two metrics, so $b(t) \not=0$. We leave aside the interpretation of this observation and its potential physical consequences 
to see whether the solutions that differ from the standard ones allow physically feasible models.

\section{Perturbative solutions}

In what follows we study infinitesimal perturbations of the classical solutions of the Einstein equations in different
scenarios like empty universe with a cosmological constant, with and without curvature,  the matter dominated (i.e. for $w=0$) 
flat universe with a vanishing cosmological constant $\Lambda_e=0$ and the radiation dominated (i.e. for $w=\frac{1}{3}$) flat 
universe.

Our working assumption is that we look for small perturbations around the symmetric, product, geometry of the form,
\begin{equation}
\begin{aligned}
&a_1(t)=a(t)+\epsilon r_1(t), \qquad &a_2(t)=a(t)+\epsilon r_2(t) ,  \qquad 
&b(t)= \epsilon s(t). 
\end{aligned}
\end{equation}
and linearise the equations of motion, taking the first terms in $\epsilon$.

In the zeroth order, we obtain (from all equations, as expected):

\begin{equation}
6 \, \frac{\left(\dot{a}(t)\right)^2}{a(t)^2} - \Lambda_e + 6\, \frac{k}{a(t)^2} = 2\, \frac{\eta}{a(t)^{3+3w}},
\label{FE}
\end{equation}
whereas the first order yields the following set of linear equations for $r_1,r_2$ and $s$, for the function $a(t)$, which already 
satisfies the equation \eqref{FE}:
\begin{equation}
\begin{aligned}
&\dot{r}_1(t)  =   \frac{3\lambda^2 a(t)^2(1+w)-(\dot{a}(t)^2+k)(1+3w)}{2a(t)\dot{a}(t)}r_1(t)+\left(\dot{a}(t) +\alpha\frac{a(t)^2}{6\dot{a}(t)}\right)s(t), \\
&\dot{r}_2(t)  =   \frac{3\lambda^2 a(t)^2(1+w)-(\dot{a}(t)^2+k)(1+3w)}{2a(t)\dot{a}(t)}r_2(t)-\left(\dot{a}(t)+\alpha\frac{a(t)^2}{6\dot{a}(t)}\right)s(t), \\
&\dot{s}(t)  =  \frac{3}{2}\frac{\dot{a}(t)}{a(t)} \left( \frac{ r_1(t)-r_2(t)}{a(t)} - 2s(t)\right),
\end{aligned}
\label{Eq1}
\end{equation} 
where we have introduced $\Lambda_e = 6 \lambda^2$ for simplicity, and denote the time derivative by a dot.

Note that for a given background solution $a(t)$ we have a homogeneous equation for the sum $r_1(t)+r_2(t)$, which has a simple
solution that, however satisfies reasonable initial conditions $r_1(t_0)=r_2(t_0)=0$ if and only if it is constantly $0$. Therefore, we may
freely restrict ourselves to the case $r_1(t) = r(t) = - r_2(t)$, and final set of perturbative equations, 

\begin{equation}
\begin{aligned}
&\dot{r}(t)  =   \frac{3\lambda^2 a(t)^2(1+w)-(\dot{a}(t)^2+k)(1+3w)}{2a(t)\dot{a}(t)}r(t)+\left(\dot{a}(t) +\alpha\frac{a(t)^2}{6\dot{a}(t)}\right)s(t), \\
&\dot{s}(t)  =   3\frac{\dot{a}(t)}{a(t)} \left(\frac{r(t)}{a(t)} - s(t)\right). 
\end{aligned}
\label{Eq2}
\end{equation} 
%%%%%%%%%%%%%%%%%%%%%%%%%%%%%%%%%%%%%%%%
%%%%%%%%%%%%%%%%%%%%%%%%%%%%%%%%%%%%%%%%

\subsection{The empty  universe}

In the case of an empty, or dark-energy dominated universe, we have the simple case of $\eta=0$ and cosmological 
solutions depending only on the curvature $k$ and the cosmological constant $\Lambda_e$.

\subsubsection{De Sitter universe ($k=0$)}
The solution of \eqref{FE} is,
\begin{equation}
a(t)=a_0 \exp\left(\sqrt{\frac{\Lambda_e}{6}}t\right), 
\label{deSitter}
\end{equation}
and the equations of motion for $r,s$ are:
\begin{equation}
\begin{aligned}
&\dot{r}(t)  =  \lambda r(t)   + a(t)  s(t) \left( \lambda + \frac{\alpha}{6 \lambda} \right), \\
&\dot{s}(t)  =  3 \lambda \frac{r(t)}{a(t)}  - 3\lambda s(t). 
\end{aligned}
\label{E1}
\end{equation}
Solving this system of linear equations we obtain, 
\begin{equation}
\begin{aligned}
s(t) = C_1 e^{-\frac{3}{2} \lambda t + \frac{1}{2} \sqrt{21 \lambda^2 + 2 \alpha} t}
   +    C_2 e^{-\frac{3}{2} \lambda t - \frac{1}{2} \sqrt{21 \lambda^2 + 2 \alpha} t}, \\
r(t) = C_3 e^{-\frac{1}{2} \lambda t + \frac{1}{2} \sqrt{21 \lambda^2 + 2 \alpha} t}
   +    C_4 e^{-\frac{1}{2} \lambda t - \frac{1}{2} \sqrt{21 \lambda^2 + 2 \alpha} t},
\end{aligned}
\end{equation}

where 
\begin{equation}
\begin{aligned}
C_3 = C_1 \frac{a_0}{6 \lambda}\,  \left( 3 \lambda  + \sqrt{21 \lambda^2 + 2 \alpha}   \right), \qquad
 C_4 = C_2  \frac{a_0}{6 \lambda}\, \left( 3 \lambda  - \sqrt{21 \lambda^2 + 2 \alpha}   \right). 
\end{aligned}
\end{equation}

\begin{figure}[h!tb]
\centering
\includegraphics[scale=1]{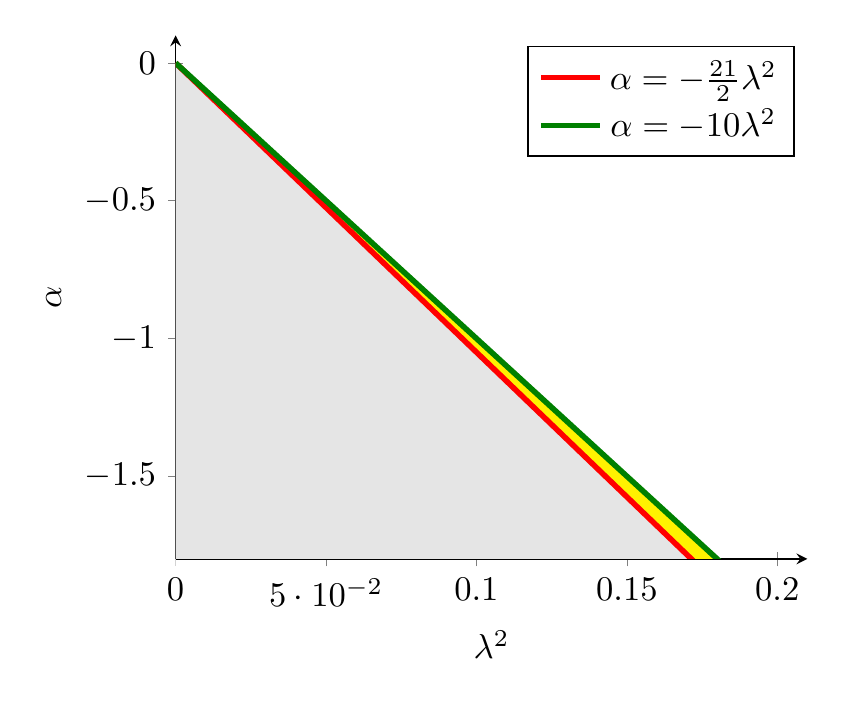}
\caption{Plot representing sectors in parameters $(\lambda^2,\alpha)$ with different behaviour of solutions.}
\label{fig:solutions_0}
\end{figure}

Depending on the relative values of the parameters $\Lambda_e=6\lambda^2$ and $\alpha$ the character of the solutions changes. For the parameters $\lambda, \alpha$, as shown on the graph on Fig. \ref{fig:solutions_0}, in the yellow region between the green and red line we have only damping exponentially decreasing solutions for $r(t)$, while in the grey region below the red line the exponentially vanishing solution is modified by oscillations.  
On the red line, however, the above form of solutions degenerates, and the correct ones are
\begin{equation}
r(t)=C e^{-\frac{1}{2}\lambda t},\qquad r(t)=Cte^{-\frac{1}{2}\lambda t}.
\end{equation}

On the other hand, we see that the perturbative solutions cannot be extended to $-\infty$ as, independently of the value of the parameters, 
they then become much bigger than the de Sitter solution. This puts the limits of applicability of the perturbative expansion which is entirely
consistent with the dark-energy dominated universe solutions. As a last remark we note that even independently of the value of $\alpha$
perturbations, which are decaying exponentially, are possible for certain values of initial parameters. For example,
if at $t=0$ we set, 
$$ r(0) =  a_0\, \left( 1 - \frac{\sqrt{21 \lambda^2 + 2 \alpha}}{3 \lambda} \right) s(0), $$
then $C_1=C_3=0$ and the perturbations will be exponentially damped for all range of parameters. 

\subsubsection{Geometries with positive and negative curvatures $k=\pm 1$}

We start with the easier case of negative curvature, for which the solution of \eqref{FE} is:
\begin{equation}
a(t)=\frac{1}{\lambda} \sinh\left( \lambda(t-t_0)\right),
\end{equation}   
and in what follows we choose $t_0=0$ to simplify the notation. 

It is convenient to change the variables and write the equations \eqref{Eq2} in $\tau=  \sinh(\lambda t)$. Then we  obtain,
\begin{equation}
\begin{aligned}
\lambda \dot{r}(\tau)   &=  \left( 1+ \frac{\alpha \tau^2}{6 \lambda^2 (1+\tau^2)} \right) s(\tau) + \frac{\lambda \tau}{1+\tau^2} r(\tau), \\
\lambda  \dot{s}(\tau)  &= 3 \frac{\lambda^2}{\tau^2} r(\tau) - \frac{3 \lambda}{\tau} s(\tau).
\end{aligned}
\label{E1k1}
\end{equation}

The above set of equations can be solved explicitly, and the solution for $s(\tau)$ is given by,
\begin{equation}
\begin{aligned}
s(t) &=  c_1 \, _2F_1\left(\frac{3}{4}-\zeta,
 \frac{3}{4}+\zeta ;3;-\tau^2\right)  
+ c_2 \, G_{2,2}^{2,0}\left( -\tau^2 \bigg|
\begin{array}{c}
\frac{1}{4} - \zeta, \frac{1}{4} + \zeta  \\
-2,0 \\
\end{array}
\right) 
\end{aligned}
\end{equation}
where $_2F_1$ is the hypergeometric function, $G_{2,2}^{2,0}$ is the generalized Meijer's function \cite{meijer} and
$$ \zeta = \frac{\sqrt{21 \lambda ^2+2 \alpha }}{4 \lambda}. $$

Since the solution is of the Big-Bang cosmology type we shall look for the small $t$ (small $\tau$) behaviour of solutions. Both functions are defined in the region $\tau^2 <1$ and can be extended analytically to the other values of $\tau^2$, 
yet $\tau^2=1$ is the point at which they are discontinuous or singular.  Additionally the Meijer's function has a pole at $0$ 
of order at least $2$ unless the parameter $\zeta$ is quantized, 
\begin{equation}
 \zeta =  \frac{9}{4} + n, \;\; n \in \mathbb{N},
\end{equation}
when it becomes regular (though non-zero). For above values of the parameter $\zeta$, the first part of the solution can be rewritten
as 
$$ c_1 \, (1+ \tau^2)^\frac{3}{2} \, {}_2F_1\left( \frac{9}{2} +n, -n ;3; -\tau^2 \right), $$
and the last component is, in fact, a polynomial of degree $n$.

The possibility of having both solutions regular at $\tau=0$ means that there exists a nonzero perturbation of the standard solution, 
which has both perturbations vanishing at the initial time $s(0)=r(0)=0$. However, the fact that $\tau=1$ is a singular point of the 
Meijer's function restricts the possibility of extending the assumed linearised perturbation beyond certain time frame. The long-time behaviour of the solutions that are arbitrary (not necessarily vanishing) at $t=0$ is  similar to the flat case and governed by value of $\zeta$, with asymptotically vanishing solutions for the same range 
of parameters $\alpha,\lambda$ as in the $k=0$ situation.

Finally, for the positive curvature, $k=1$, the pure dark energy solution is, 
\begin{equation}
a(t)=\frac{1}{\lambda} \cosh\left( \lambda(t-t_0)\right), 
\end{equation}   
and the small perturbations at $t_0=0$ are, again changing the variable to $\tau=\sinh(\lambda t)$,

\begin{equation}
\begin{aligned}
\lambda \dot{r}(\tau)   &=  \frac{1}{\sqrt{1+\tau^2}} \left( \tau + \frac{\alpha (1+\tau^2)}{6 \lambda^2 \tau} \right) s(\tau) + \frac{\lambda}{\tau} r(\tau), \\
\lambda  \dot{s}(\tau)  &=  \frac{3}{\sqrt{1+\tau^2}}  \frac{\lambda^2 \tau}{1+ \tau^2} r(\tau) - \frac{3 \lambda \tau}{1+\tau^2} s(\tau).
\end{aligned}
\label{E2k2}
\end{equation}
which, similarly as in the previous situation, has the solutions that are expressed in terms of the hypergeometric function ${}_2F_1$:
\begin{equation}
s(t)=c_1\, {}_2F_1 \left(\frac{3}{4}-\zeta, \frac{3}{4}+\zeta;-\frac{1}{2};-\tau^2\right)+c_2 \, t^3 {}_2F_1\left(\frac{9}{4}-\zeta,\frac{9}{4}+\zeta;\frac{5}{2};-\tau^2\right).
\end{equation} 
From the fact that in this case
$$r(\tau)\sim -\frac{\alpha c_1}{6\lambda^3}+\frac{c_2 \tau }{\lambda}+O(\tau^2),$$
we deduce that if we require $r(0)=0$ then $c_1=0$.  One can easily check that then also $s(0)=0$, however, both solutions will grow with $t$. On the other
hand, the exponentially decreasing solution requires $c_2=0$. 
%%%%%%%%%%%%%%%%%%%%%%%%%%%%%%%%%%%%%%%%%%%%%%%%%%%%%%%%%%%%%%%%%%
\subsection{Matter dominated universe}

In a completely similar manner we consider the limit in a matter-dominated universe, in which 
we put $\Lambda_e=0$ and $w=0$, while $\eta\not=0$.  We start with the  Einstein-de Sitter universe, $k=0$. 
In this case  the standard solution,
$$ a(t) =  \left( \frac{3}{4} \eta \right)^{\frac{1}{3}} t^\frac{2}{3}, $$
gives the following equations for $r(t)$ and $s(t)$,
\begin{equation}
\begin{aligned}
\dot{r}(t) &= -\frac{1}{2} \frac{ \dot{a}(t)}{a(t)}  r(t) +  \left( \frac{\alpha}{6}  \frac{a(t)^2}{\dot{a}(t)} +  \dot{a}(t) \right) s(t),  \\ 
\dot{s}(t)&= 3 \frac{\dot{a}(t)}{a(t)^2} r(t) - 3 \frac{\dot{a}(t)}{a(t)} s(t).
\label{matter}
\end{aligned}
\end{equation}
The general solution for $s(t)$ can be expressed in terms of  Bessel functions,
\begin{equation}
s(t)=c_1 t^{-\frac{3}{2}} J_{\sqrt{\frac{19}{12} }} \left(\sqrt{-\frac{\alpha}{2}}\, t\right) + 
       c_2  t^{-\frac{3}{2}} Y_{\sqrt{\frac{19}{12}}} \left(\sqrt{-\frac{\alpha}{2}}\, t\right) ,
\end{equation}
and the solution for $r(t)$ can be, consequently derived from the second of \eqref{matter}. In case of negative $\alpha$ 
the long-time solutions have oscillatory character with the following asymptotic behaviour of their amplitudes: 
$$ s(t) \sim t^{-2}, \qquad r(t) \sim t^{-\frac{1}{3}}, $$		
so for $\alpha<0$ the perturbations decay in $t$ independently of the initial values of the perturbation at any fixed time. Although
the matter-dominated universe describes rather later periods in the evolution of the universe, still there exists a solution, which 
is regular at $t=0$. 

For positive values of $\alpha$ only the second solution, which is exponentially decaying, is an acceptable one as a perturbation, which signifies that
for this range of the parameter only specific perturbations are stable.

\subsection{Radiation dominated universe}
For this situation (again $\Lambda_e=0, k=0$) the standard solution of the Einstein equations is,
$$ a(t) =  \left( \frac{4}{3} \eta \right)^{\frac{1}{4}} t^\frac{1}{2}, $$
which gives us the following equations for the perturbations:

\begin{equation}
	\begin{aligned}
		\dot{r}(t) &= - \frac{ \dot{a}(t)}{a(t)}  r(t) +  \left( \frac{\alpha}{6}  \frac{a(t)^2}{\dot{a}(t)} +  \dot{a}(t) \right) s(t),  \\ 
		\dot{s}(t)&= 3 \frac{\dot{a}(t)}{a(t)^2} r(t) - 3 \frac{\dot{a}(t)}{a(t)} s(t).
	\end{aligned}
\end{equation}

The solutions for $s(t)$ is, 
\begin{equation}
	s(t)=  c_1 t^{-\frac{5}{4}} J_{\sqrt{\frac{13}{16} }}\left(\sqrt{-\frac{\alpha}{2}}\,t\right) 
	     +  c_2  t^{-\frac{5}{4}} Y_{\sqrt{\frac{13}{16}}}\left(\sqrt{-\frac{\alpha}{2}}\, t\right) ,
\end{equation}
with the exact expression for $r(t)$ that can be obtained directly from the second equation. 

Again, in the case of $\alpha<0$ the long-time behaviour of the amplitude of oscillations is
$$ s(t) \sim t^{-\frac{7}{4}}, \qquad r(t) \sim t^{-\frac{1}{4}}.$$		

However, a very interesting situation occurs near the Big Bang, $t=0$, as in the best case the solution for $s(t)$
diverges and behaves like $t^{\frac{\sqrt{13}-5}{4}}$, whereas the scale factor $r(t)$ behaves like $t^{\frac{\sqrt{13}-3}{4}}$
and is regular. The same result will be valid for $k=\pm 1$, as the near Big Bang asymptotics of the radiation dominated
universe has the same structure. 

The explicit solutions for the $k=-1$ geometry are in terms of the confluent Heun functions and the long-time dependence of 
the perturbations will be again similar for $\alpha<0$ as is suggested by a brief numerical analysis of example solutions.  

As the solutions for $k=1$ are cyclic, the long-term asymptotic of the perturbations does not make sense in this case.

\section{Summary and outlook}

The simplest almost-commutative geometry of the two-sheeted universe, which is motivated by the Connes-Lott idea \cite{CoLo} 
is an interesting model to study its potential relevance not only for the particle physics but also for its implication to the large-scale structure
of the Universe. We have shown that an abstract model, with a more general type of metric structure that is not necessarily a product
structure allows a two-metric theory, which is very similar to the bimetric theory of gravity. Although we are aware that both the
interaction structure as well as the interpretation of the model's origin are quite different there are striking similarities in the potential
term of the action. It shall be noted that models \bblu originating \eblu from quantum deformations of spacetime have a similar feature of
two metrics although their origins are different \cite{CSV19}.

Leaving the full model that was developed for the particle interactions \cite{connes1, connes2} aside and concentrating first on a simplified
one, we have focused on a primary question of stability of classical Friedmann-Lema{\^i}tre-Robertson-Walker solutions. 
To be more precise, our idea was to check whether for some range of parameters a small perturbation in the Dirac operators making
\bblu the full one\eblu, and hence the metrics different from each other on the two sheets of the Universe, will diverge or collapse.

Our conclusion is that for the considered range of models, including flat and curved spatial geometries with dark-energy, radiation
or matter dominance there exist a range of parameters so that the symmetric solution (product geometry) is dynamically stable. Our 
analysis confirms but hugely extends the earlier indications \cite{sitarz2019} by allowing both the scale factors as well as lapse 
functions to vary. The stability of the cosmological solutions suggests that the models with two metrics are admissible from the 
physical point of view and are an interesting modification of geometry that may be used in future models.

This has an important bearing on the physical consequences of the model. First of all, cosmological observables like redshift 
and observable Hubble constant will be related to the background standard Friedmann-Lema{\^i}tre-Robertson-Walker solution. 
This follows from the fact that both light and matter will couple (as argued in section \ref{s3}) to both metrics and, taking into account 
that in most models the difference between metrics is decreasing as the Universe evolves, only the average (background) scale factor 
$a(t)$ will determine the observable redshift. However, one can speculate that a possible sign of the fluctuating two metrics might be seen 
in physical effects that couple only to one metric (as might be \bblu the \eblu case of massless Majorana particles) or couple to metrics \bblu in \eblu a nonlinear way. 

The constructed (simplified) model is predominantly based on the idea that allowed to explain the appearance of Higgs field and Higgs
quartic symmetric-breaking potential from purely geometric considerations as a form of generalized gauge theory. Transferring this 
concept to the theory of metric and generalized general relativity appears to be a natural an well-motivated physical step. Unlike in the 
bimetric theory, here the interaction terms between the two metrics are completely determined by the structure of the theory yet are not 
computable in full generality. This prevents us from an analysis of the possibility of ghost-free sectors in the way it was done for bimetric
theories \cite{hr2,kl} . Nevertheless, since the model has strong features similar to bimetric gravity  (as we have stressed in \ref{sG}), in particular,
even though the effective interaction potential between the two metric is not a symmetric polynomial of $\sqrt{g_1^{-1} g_2}$ but rather 
a rational function, where the nominator and denominator are of this form, we expect that a similar result will hold. 

Apart from the fundamental questions of physical consistency and interpretation of the degrees of freedom of the theory there are still 
several questions that remain open. First of all, in case of small deviations from the product geometry it is interesting whether they might 
have some observable physical consequences both in the pure gravity sector as well as in the sector of the matter and radiation. 
Though this might be considered as pure speculation, such  fluctuations of the metrics, if existing in the radiation era, might be linked to some parity 
anisotropies \cite{anis} in the Cosmic Microwave Background radiation. Another possible sector of the theory to explore are solutions with 
singularities like black holes. All such ideas need to be explored carefully in future studies.

\section{Acknowledgements} 
AB acknowledges the support from the National Science Centre, Poland, ~grant 2018/31/N/ST2/00701.

\appendix
\section{Symbols of the operator $D^{-2}$}\label{sec:symbols}
Suppose $P$ and $Q$ are two pseudodifferential operators with symbols
\begin{equation}
\sigma_P(x,\xi)=\sum\limits_{\alpha} \sigma_{P,\alpha}(x)\xi^\alpha, \hspace{20pt} \sigma_Q(x,\xi)=\sum\limits_{\beta} \sigma_{Q,\beta}(x)\xi^\beta,
\end{equation}
respectively, where $\alpha, \beta$ are multiindices. The composition rule takes the following form \cite{gilkey}
\begin{equation}
\sigma_{PQ}(x,\xi)=\sum\limits_{\gamma}\frac{(-i)^{|\gamma|}}{\gamma!}\partial^\xi_\gamma \sigma_{P}(x,\xi)\partial_\gamma \sigma_Q(x,\xi),
\label{composition}
\end{equation}  
where $\partial_a^\xi$ denotes partial derivative with respect to coordinate of the cotangent bundle.

Let us consider the case when $P=D^{-2}$ and $Q=D^2$. Since $D^2$ has a symbol
\begin{equation}
\sigma_{D^2}(x,\xi)=\mathfrak a_2+\mathfrak a_1+\mathfrak a_0,
\end{equation}
then $D^{-2}$ has to have a symbol of the form
\begin{equation}
\sigma_{D^{-2}}(x,\xi)=\mathfrak b_0+\mathfrak b_1+\mathfrak b_2+...,
\end{equation}
where $\mathfrak b_k$ is homogeneous of order $-2-k$.

Inserting these expressions into \eqref{composition} and taking homogeneous parts of order $0,-1$ and $-2$ we get the following set of equations:
\begin{equation}
\begin{aligned}	
&\mathfrak b_0 \mathfrak a_2=1, \\
&\mathfrak b_0 \mathfrak a_1+\mathfrak b_1\mathfrak a_2-i\partial_a^\xi(\mathfrak b_0)\partial_a(\mathfrak a_2)=0, \\
&\mathfrak b_2\mathfrak a_2+\mathfrak b_1\mathfrak a_1+\mathfrak b_0\mathfrak a_0-i\partial_a^\xi(\mathfrak b_0)\partial_a(\mathfrak a_1)-i\partial_a^\xi(\mathfrak b_1) \partial_a(\mathfrak a_2)-\frac{1}{2}\partial_a^\xi\partial_b^\xi(\mathfrak b_0)\partial_a\partial_b(\mathfrak a_2)=0,
\end{aligned}
\end{equation}
From these relations we get
\begin{equation}
\begin{aligned}
&\mathfrak b_0=\mathfrak a_2^{-1}, \\
& \mathfrak b_1=- \left(\mathfrak b_0 \mathfrak a_1-i\partial_a^\xi(\mathfrak b_0)\partial_a(\mathfrak a_2)\right)\mathfrak b_0,\\
&\mathfrak b_2= -\left(\mathfrak b_1\mathfrak a_1+\mathfrak b_0\mathfrak a_0-i\partial_a^\xi(\mathfrak b_0)\partial_a(\mathfrak a_1)-i\partial_a^\xi(\mathfrak b_1) \partial_a(\mathfrak a_2)-\frac{1}{2}\partial_a^\xi\partial_b^\xi(\mathfrak b_0)\partial_a\partial_b(\mathfrak a_2)\right)\mathfrak b_0.
\end{aligned}
\end{equation}

%%%%%%%%%%%%%%%%%%%%%%%%%
%%%%%%%%%%%%%%%%%%%%%%%%%

\end{document}